\documentstyle[bezier,12pt]{article}

\newcommand{\bra}[1]{\langle #1 |}
\newcommand{\ket}[1]{| #1 \rangle}

\newcommand{\eq}[1]{eq.(\ref{#1})}

\newcommand{\ddpar}[2]{\frac{\partial^2 #1}{\partial #2^2}}

\newcommand{\bm}[1]{\mbox{\boldmath $#1$}}

\def\ghost#1{\vrule height#1 depth#1 width0pt \displaystyle}
\def\const{\mbox{const}}
\def\e{\mbox{e}}

\def\sn{\mbox{sn}}

\def\sn{\mbox{sn}}

\def\Im{\mbox{Im}}
\def\dn{\mbox{dn}}

\begin{document}
\title{Examples of Semiclassical Instanton-Like Scattering: Massless
$\phi^4$ and SU(2) Gauge Theories}
\author{
  D.T.Son and P.G.Tinyakov \\
  {\small \em Institute for Nuclear Research of the Russian Academy of
  Sciences,}\\
  {\small \em 60th October Anniversary prospect 7a, Moscow 117312}
  }
\maketitle

\begin{abstract}
Two-parameter sets of solutions to the classical field equations in the
massless $\phi^4$ model and SU(2) gauge theory are found, each solution
presumably describing a multi-particle instanton-like transition at high
energy.  In the limit of small number of initial particles, the
probability of the transition is suppressed by $\exp(-2S_0)$, where
$S_0$ is the instanton action.
\end{abstract}

\newpage

\section{Introduction}

     For quite a while, there has been a considerable interest in the
problem of non-perturbative scattering at parametrically high energies in
weakly coupled theories. In the context of the standard model, this
problem is usually associated with the anomalous baryon number violation
which might be observable in particle collisions in 10-100 TeV region.
Despite the considerable effort made since the first quantitative attempts
to estimate the corresponding cross section \cite{Ringwald90,Espinosa90},
the complete understanding of this problem has not yet been achieved (for a
review of the present status see refs.\cite{Mattis-rev,T-rev}).
Although the total cross section at high energies has been strongly
argued \cite{Mueller-92} to have the semiclassical form
  \begin{equation}
  \sigma(E)=\exp \biggl\{ {1\over g^2}F\Bigl({E\over E_0}\Bigr) \biggr\} \; ,
  \label{sigma of E}
  \end{equation}
where $g$ is the coupling constant and $E_0\sim M_W/g^2$ is the
characteristic energy scale (the sphaleron scale), the explicit procedure
for evaluating the function $F(E/E_0)$ is still unknown, the main difficulty
consisting in the semiclassical treatment of high energy initial particles.

To overcome this difficulty, there has been recently proposed an implicit
approach \cite{RT,T} which makes use of the idea to approximate the
two-particle cross section in the one-instanton sector by the probability of
a multiparticle instanton-like transition for a small number of initial
particles. If the number of initial particles scales like $n=\nu/g^2$, where
$\nu$ is a (small) numerical constant, the latter probability is explicitly
semiclassical,
  \begin{equation}
  \sigma(E,n)=\exp \biggl\{ {1\over g^2}F\Bigl({E\over E_0},\nu\Bigr)
  \biggr\} \; ,
  \label{sigma of E,n}
  \end{equation}
while the function $F(E/E_0,\nu)$ is calculable by semiclassical methods
\cite{RT,T,RST1} for any particular coherent initial state
containing $n=\nu/g^2$ particles.

     There are two possible ways of extracting information about the
two-particle cross section from this sort of considerations. First, one may
study the probability of the transition summed over (or, equivalently,
maximized with respect to) some subset of fixed-energy initial states. The
obvious, but by no means unique choice could be the set of all $n$-particle
states of fixed energy. This maximum probability is again semiclassical,
i.e. has the form (\eq{sigma of E,n}), where the exponent depends, of
course, on the particular choice of the subset of the initial states. By
construction, for all sets that include the particular state where two
particles carry almost all the energy, the corresponding function
$F(E/E_0,\nu)$ sets an {\it upper bound\/} for the logarithm of the
two-particle cross section, $F(E/E_0)$. The advantage of this way is that
it allows not to deal with (probably irrelevant) details of the initial
state.

An alternative way is to consider the transition from some particular
coherent state $\ket{a}$ containing $n=\nu/g^2$ particles. The perturbative
calculations in three first orders in the electroweak theory indicate that,
in the limit $\nu\to 0$, the corresponding function $F_a(E/E_0,\nu)$ is
smooth and independent of the details of the initial state
\cite{Mueller-92'p}. If this property held in higher orders, the limiting
value $F_a(E/E_0,\nu\to 0)$ would coincide with the small $\nu$ limit of the
maximum probability discussed above and, furthermore, would coincide with
the logarithm of the two-particle cross section, $F(E/E_0)$.

     Clearly, the first way is preferable as it gives the upper bound on the
two-particle cross section regardless of whether the assumption about the
independence of the probability of the precise form of the initial state
is justified. Unfortunately, it requires solving the field
equations with particular boundary conditions \cite{RST1}, which is
almost certainly beyond the scope of analytical methods. The more realistic
way could be to find {\it any\/} solutions to the field equations and
then check if they correspond to some instanton-like transition with small
number of initial particles. This is our strategy in the present paper.

There are two requirements which necessarily have to be satisfied by the
solution saturating the total probability of instanton-like transition
from some initial state \cite{RST1}:
\begin{enumerate}
\item[i)]
The solution must be instanton-like, i.e. must interpolate between two
different asymptotic regions separated by a potential barrier. The
underbarrier part of the field configuration is described by the evolution
in the imaginary time (the ``duration'' of the Euclidean stage $T$ is
infinite for the instanton and finite for finite energy transitions). So,
the solution we are interested in should exist on the contour ABCD of
fig.1 in the complex time plane and satisfy the complexified field
equations along this contour. Since the initial and final asymptotic
regions, A and D in fig.1, must lie on different sides of the barrier,
there must be singularities between the line AB and the negative part
of the real time axis, which prevent from doing trivial analytic
continuation between these two regions.
\item[ii)]
The solution must be real on the real time axis. This requirement is
valid as long as the probability summed over the final states is
considered. Note that it can be viewed as one of the two boundary
conditions necessary to specify a unique solution to the field equations.
In what follows we will impose a stronger condition, namely requirement
that the solution has a turning point at $t=0$, i.e. is also real at the
imaginary time axis.
\end{enumerate}
Any solution satisfying these two conditions corresponds, in general, to
the instanton-like transition from some initial state to the state on
the other side of the barrier which maximizes the transition probability.
In other words, the solutions of this type saturate the total probability
  \begin{equation}
  \sigma_a(E) \sim \sum_b|\bra{b}SP_E\ket{a}|^2
  \label{sigma a}
  \end{equation}
at some initial state $\ket{a}$. Here $P_E$ is the projector onto fixed
energy $E$.

Given the solution satisfying the conditions i) and ii), one can easily
find the characteristics of the instanton-like transition it corresponds
to.  Assuming that fields are free in the asymptotic regions A and D, one
writes
  \begin{eqnarray}
  \phi_{i}(\bm{k},\eta) & = & \ghost{15pt}
  {(2\pi)^{3/2}\over \sqrt{2\omega_{\bf k}}}
  ( f_{\bf k}\e^{-i\omega_k\eta} + \bar{f}_{-{\bf k}}
  \e^{i\omega_k\eta}) \; , \label{phi = fe + fe} \\
  \phi_f(\bm{k},t) & = &\ghost{15pt}
  {(2\pi)^{3/2}\over \sqrt{2\omega_{\bf k}}}
  ( b_{\bf k}\e^{-i\omega_kt} + b^*_{-{\bf k}}
  \e^{i\omega_kt})  \; ,
  \label{phi = ge + ge}
  \end{eqnarray}
where $\eta=t-iT$ is a real parameter along the line AB. Note that because
of the condition ii) the amplitudes $b_{\bf k}$ and $b^*_{\bf k}$ are
complex conjugate to each other, while $f_{\bf k}$ and $\bar{f}_{\bf k}$
are not. The most probable final state of the transition can be read off
from \eq{phi = ge + ge}. This is the coherent state defined by the complex
variables $b_{\bf k}$, $\ket{b}= \exp(\int\!d\bm{k} b_{\bf k}
\hat{b}^{\dagger}_{\bf k}) \ket{0}$. Correspondingly, the number of
particles in the final state and its energy are
  \begin{equation}
  \begin{array}{rcl}
  n_f  & = & \ghost{15pt} \int\!d\bm{k}b^*_{\bf k}b_{\bf k}\; , \\
  E  & = & \ghost{15pt} \int\!d\bm{k}\omega_{\bf k}b^*_{\bf k}b_{\bf k}\;.
  \end{array}
  \label{E,n fin: def}
  \end{equation}

The complex variables characterizing the initial coherent state $\ket{a}$
can be obtained from the amplitudes $f_{\bf k}$ entering \eq{phi = fe +
fe} as follows,
  \begin{equation}
  f_{\bf k}=a_{\bf k}\e^{\omega_k\Delta T} \; .
  \label{ f_k = a_k }
  \end{equation}
The presence of the additional parameter $\Delta T$ reflects the fact that
the coherent states $\ket{a_{\bf k}}$ and $P_E\ket{a_{\bf k}}$ differ only
by normalization \cite{RST2p}. The number of particles in the initial
state depends only on the difference $T-\Delta T$,
  \begin{equation}
  n_i = \int\!d\bm{k}a^*_{\bf k}a_{\bf k}
  = \int\!d\bm{k}|f_{\bf k}|^2 \e^{-2 \omega_k\Delta T} \; .
  \label{n in: def}
  \end{equation}
The latter difference can be found from the energy conservation,
  \begin{equation}
  \int\!d\bm{k} \omega_{\bf k} b_{\bf k} b^*_{\bf k}
  = \int\!d\bm{k} \omega_{\bf k} a_{\bf k} a^*_{\bf k}
  \equiv \int\!d\bm{k}\omega_{\bf k} |f_{\bf k}|^2
  \e^{-2 \omega_k\Delta T}  \; .
  \label{E=aa*}
  \end{equation}
Finally, the probability of the transition is given by the formula
  \begin{equation}
  \sigma=\exp\left\{ 2E(T-\Delta T)
  -2\mbox{Im}~S(\phi) + \int\!d\bm{k}(\bar{f}_{\bf k} f_{\bf k}
  - a^*_{\bf k}a_{\bf k}) \right\} \; ,
  \label{probability}
  \end{equation}
where $S(\phi)$ is the action calculated along the contour ABCD of fig.1.

The examples of the solutions satisfying  the requirements i) and ii) were
recently found in the two-dimensional O(3) $\sigma$-model \cite{RST2p}. In
this paper we generalize the method of ref.\cite{RST2p} to the
four-dimensional case, namely, to the massless $\phi^4$ model and the
massless SU(2) gauge theory. The latter can be viewed as the high energy
limit of the electroweak theory. In sections 2 and 3 devoted to the
$\phi^4$ model and SU(2) theory, respectively, we show that in both cases
there exists a two-parameter family of the solutions satisfying the
requirements i) and ii). One of these parameters is a trivial scale
parameter which is due to the conformal symmetry of the models. Another
parameter, $\epsilon$, is dimensionless. At small values of this
parameter the corresponding instanton-like transition is characterized
by the number of initial particles much smaller than the number of final
particles, both being much less than $1/g^2$, the characteristic number
of particles in the sphaleron solution. The probability of the transition
calculated by means of \eq{probability} is $\exp (-2S_{inst}
+O(\epsilon^2) )$, i.e. is suppressed by the same factor as at zero
energies. The existence of such solutions may indicate that the
probability of the instanton-like transitions from the initial states
containing small number of particles is suppressed at high energies by the
same suppression factor as at zero energy. Further discussion of this
possibility is contained in sect.4, where we also present our conclusions.

\section{$\phi^4$ model}

As a simplest four-dimensional model, consider the $\phi^4$
theory with negative coupling constant. This model was suggested as a
suitable testing ground for studying instanton transition at high
energies \cite{Voloshin-91,Hsu}. The Euclidean Lagrangian of the model
reads
  \begin{equation}
  L= {1\over 2}(\partial_{\mu} \phi)^2 - {\lambda\over 4!}  \phi^4 \; ,
  \label{phi4lagr}
  \end{equation}
where $\lambda$ is positive.
In this model, the state $\phi=0$ is metastable with respect to
non-homogeneous fluctuations. The decay of this state is described by
the instanton solution
  \begin{equation}
  \phi_{inst}(x) = {2\sqrt{3}\over\sqrt{\lambda}}{\rho\over
  x^2+\rho^2} \; .
  \label{phi inst}
  \end{equation}
The action of the instanton is
  \[
  S= {16\pi^2\over \lambda}
  \]
Following refs.\cite{Voloshin-91,Hsu}, one may consider the scattering of
particles in the background of the instanton solution $\phi_{inst}$.
These processes share many common features with instanton-like
transitions in the electroweak theory (such as exponential growth
of the total cross section at small energies) and formally may be used
as a laboratory for studying the instanton effects in particle
collisions.

As discussed in the Introduction, the probability of the instanton-like
transition is saturated by the classical solutions satisfying the
requirements i) and ii). To find such solutions one can use the method of
ref.\cite{RST2p} based on the conformal invariance. First, one looks for an
$O(4)$ symmetric solution to the Euclidean field equations. Making use of
the spherically symmetric ansatz
  \begin{equation}
  \phi(y_\mu)=\frac{2\sqrt{3}}{\sqrt{\lambda}} \frac{1}{y} f(\xi) \; ,
  \label{sphsymsol}
  \end{equation}
where $\xi=\ln(y/R)$ and $R$ is a parameter with the dimension of length, we
find the equation for $f(\xi)$ which is formally the classical
equation of motion for a particle moving in one-dimensional double-well
potential,
  \[
  \ddpar{f}{\xi} = f-2f^3 \; .
  \]
The solution to this equation with the turning point at $\xi=0$ reads
  \begin{equation}
  f(\xi)=\frac{1}{\sqrt{1+\epsilon^2}}\dn\left(
  \frac{\xi}{\sqrt{1+\epsilon^2}} +  K(k); k\right) \; ,
  \label{solforf}
  \end{equation}
where $k=(1-\epsilon^2)^{1/2}$ and $\epsilon$ is a remaining free
parameter which varies in the range $0\leq \epsilon \leq 1$. In \eq{solforf}
$K(k)$ is the complete elliptic integral. The instanton solution
(\ref{phi inst}) is reproduced at $\epsilon=0$ ($k=1$).

In order to obtain the solution to the field equations which has the
turning point at $t=0$, we perform the conformal transformation to the
new variables $x_{\mu}$,
  \begin{equation}
  x_{\mu}=\frac{2R^2}{(y+a)^2}(y_\mu+a_\mu)-a_\mu     \; ,
  \label{conftrans}
  \end{equation}
with $a_\mu=(R,0,0,0)$, which maps the sphere $y^2=R^2$
onto the upper half of the $x$-space. The solution (\ref{sphsymsol})
becomes
  \[
  \phi_E(x)=\frac{4\sqrt{3}}{\sqrt{\lambda}}\frac{R f(\xi)}{\sqrt{((x_0
  -R)^2+r^2)((x_0+R)^2+r^2)}}  \; ,
  \]
where $r=|\bm{x}|$.
It is convenient to perform the analytical continuation to the
Min\-kows\-kian domain in this equation. Finally we obtain
  \begin{equation}
  \phi(x)=\frac{4\sqrt{3}}{\sqrt{\lambda}}\frac{Rf(\xi)}
	  {\sqrt{((t-iR)^2-r^2)((t+iR)^2-r^2)}}  \; ,
  \label{solution}
  \end{equation}
where
  \[
  \xi=\frac{1}{2}\ln\frac{(t-iR)^2-r^2}{(t+iR)^2-r^2} \; .
  \]
Due to the conformal invariance of the $\phi^4$ model, the field
configuration (\ref{solution}) is again a solution to the field
equations. Furthermore, since $f(\xi)$ has the turning point at
$\xi=0$, the field (\ref{solution}) is clearly symmetric under the
time reversal $t\to -t$ and, therefore, has a turning point at
$t=0$.

Consider now the structure of the solution (\ref{solution}) in the
complex time plane. As can be seen
from \eq{solution}, for each $r$ there exists an infinite set
of singularities located as shown in fig.2. At $r\to \infty$
and small $\epsilon$ the positions of these singularities can be found
analytically,
  \begin{equation}
  t_n= -r + iR\left(1-2\left(\frac{\epsilon}{4}\right)^{2(2n+1)}
       \right) \; .
  \label{poles}
  \end{equation}
It can also be shown that there exists a contour $\Im\, t = T$ such that
at all $r$ it passes above the first singularity in this
series and below the next one. Since our solution is real on the real
time axis and has a turning point, both requirements i) and ii) are
satisfied.

Now we have to determine what kind of scattering process the above
solution corresponds to. For this purpose we consider the asymptotics
of the field (\ref{solution}) in the regions $t\to
iT-\infty$ and $t\to\infty$, which correspond to the initial and final
states, respectively. In the asymptotic regions the solution becomes a linear
superposition of the plane waves, the coefficients in this
superposition determining the initial coherent state and the most
probable final state of the process. Making use of the
explicit solution (\ref{solution}), it is easy to check that in the
asymptotic regions the field is concentrated at $r -| t|\sim 1$:
$\phi$ falls rapidly as one moves off the light cone and becomes
negligibly small when $r-|t|\gg 1$. The magnitude of the field on
the light cone decreases like $1/t$ for large $t$, so that the energy
is conserved. Thus,
  \begin{equation}
  \phi_{i,f}= \frac{2\sqrt{3}}{\sqrt{\lambda}}
  \frac{1}{|t|}F_{i,f}(\delta)+O(t^{-2})  \; ,
  \label{phi=F}
  \end{equation}
where $\delta=r\pm t$ for the initial and final asymptotics, respectively.
In the asymptotic regions, the spatial Fourier transform of \eq{phi=F} reads
  \begin{equation}
  \phi_{i,f}(\bm{k}) = {4\pi\sqrt{3}\over ik\sqrt{\lambda} }
  \int\limits_{C_{i,f}}
  \!d\delta \Bigl(\e^{\mp ikt+ik\delta}
  -\e^{\pm ikt-ik\delta} \Bigr)  F_{i,f}(\delta) \; ,
  \label{fourier phi}
  \end{equation}
where $k=|\bm{k}|$. The integration contour is $\Im\,\delta=0$ for the
final state and $\Im\,\delta=T$ for the initial state.

Consider first the final state. Expanding in $\epsilon$, we find
  \[
  F_f(\delta)=\frac{\epsilon\delta}{\delta^2+1} \; ,
  \]
where $\delta=r-t$.
The spatial Fourier transform of this field at $t\to \infty$ equals
  \[
  \phi(\bm{k}) = {(2\pi)^{3/2}\over\sqrt{2\omega_{\bf k}}} \Bigl\{
  \e^{-i\omega_kt} b_{\bf k} + b^*_{\bf -k} \e^{i\omega_kt} \Bigr\} \; ,
  \]
where
  \[
  b_{\bf k}=b_{\bf k}^*=
  \frac{2\sqrt{3\pi}\epsilon R}{\sqrt{\lambda k}}\e^{-kR} \; .
  \]
These amplitudes define the final coherent state, $\exp ( \int
d\bm{k} b_{\bf k} \hat{b}_{\bf k}^{\dagger})\ket{0}$, which has the
number of particles
  \[
  n_f = \int\! d\bm{k} b^*_{\bf k} b_{\bf k} = {12\pi^2\over
  \lambda} \epsilon^2 \; ,
  \]
and the energy
  \begin{equation}
  E = \int\! d\bm{k} \omega_{\bf k} b^*_{\bf k} b_{\bf k} =
  {12\pi^2\over R\lambda} \epsilon^2 \; .
  \label{energy}
  \end{equation}

At $t\to -\infty$, we can find the function $F_i(\delta)$ in the
limit $\epsilon\to0$. In order to find the initial state, we need the
amplitudes $f_{\bf k}$ (see \eq{phi = fe + fe}). As follows from
\eq{fourier phi}, this coefficient is determined by that part of the
function $F_i(\delta)$ which has singularities above the contour
$\Im\,\delta=T$.  The lowest in $\epsilon$ contribution of this type is
  \[
  F_i(\delta)= \frac{3\epsilon^3}{16}
  \frac{\delta/R-i-i\epsilon^2/8}{(\delta/R-i+i\epsilon^2/8)^2}
  \ln(\delta/R -i) - {i\epsilon\over 4} {(\delta/R -i-i\epsilon^2/8)^2
  \over (\delta/R-i)(\delta/R-i + i\epsilon^2/8 )}
  \]
  \[
 +\;\mbox{terms, regular at $\mbox{Im}~\delta>T$}\; ,
  \]
where $\delta=r+t$.
Calculating the spatial Fourier transform by means of
\eq{fourier phi} we find
  \[
  f_{\bf k} = {\sqrt{3\pi}\epsilon^3 R\over 8\sqrt{\lambda k}}
  \biggl\{ 6 \int\limits^{\infty}_0\!du \e^{-kRu}
  {u -\epsilon^2/8\over (u +\epsilon^2/8)^2} +1 \biggr\} \e^{-k(R-T)} \; .
  \]
Eqs.(\ref{ f_k = a_k }), (\ref{E=aa*})  and (\ref{energy}) give
  \[
  T-\Delta T = R\Bigl\{ 1- \Bigl({\epsilon^2\over 4} \ln
  {1\over \epsilon}\Bigr)^{2/3}\Bigr\}\; ,
  \]
where terms subleading in $\epsilon$ are omitted.
Making use of these equations we obtain
  \begin{equation}
  n_i =  \int\! d\bm{k} a^*_{\bf k} a_{\bf k} =
  {3\cdot2^{2/3}\pi^2\over\lambda}
  \epsilon^{10/3}\ln^{2/3}\frac{1}{\epsilon}\;.  \label{n in}
  \end{equation}
Note that at small $\epsilon$ the number of initial particles is much
smaller than the number of the final ones.

The probability of the transition described by the solution
(\ref{solution}) is given by \eq{probability}. To calculate the
imaginary part of the action we note that the only contribution comes
from the singularity lying between the integration contour and the
real time axis. Thus, the time integral is reduced to the residue at
the pole. The remaining space integral can be done exactly. After some
algebra one finds
  \[
  \Im S = S_{inst} = 16\pi^2/\lambda
  \]
and, therefore,
  \begin{equation}
  \sigma= \exp\Bigl( -{32\pi^2\over\lambda} + O(\epsilon^2) \Bigr) \; .
  \label{phi4 probability}
  \end{equation}
At small $\epsilon$ the probability of the transition is exponentially
suppressed by the same factor as at zero energy.

\section{SU(2) model}

In the case of the pure SU(2) theory our strategy of obtaining the
solutions is the same as in sect.2. First, we write the spherically
symmetric ansatz
  \[
   A^a_{\mu}(y)=\frac{2}{g}\eta^{a\mu\nu}y_{\nu}\frac{f(\xi)}{y^2} \; ,
  \]
where $\eta^{a\mu\nu}$ is the t'Hooft symbol and $\xi=1/2 \ln (y^2/R^2)$.
The Yang-Mills equations, in terms of the function $f(\xi)$, read:
  \[
  -\ddpar{f}{\xi} + 4(f^2-f)(2f-1) =0 \; .
  \]
The solution with the turning point at $\xi=0$ has the form
  \[
  f=\frac{1}{2} \left(1 + k \sqrt{\frac{2}{1+k^2}} \sn \left(
    \sqrt{\frac{2}{1+k^2}} \xi-K;k \right) \right) \; ,
  \]
where $k=\sqrt{1-\epsilon^2}$ is a free parameter. In
the limit $\epsilon\to 0$ the usual instanton solution is reproduced. After
the conformal transformation (\ref{conftrans}) and analytic continuation to
the Minkowskian domain we obtain\footnote{In fact, this solution was
previously obtained in refs.\cite{Luscher,Schechter}.}
  \[
  A^a_0=-\frac{8 R t
  x_af(\xi)}{g[(t-iR)^2-r^2][(t+iR)^2-r^2]} \; ,
  \]
  \begin{equation}
  A^a_i=\frac{4R(\delta_{ai}(R^2 +t^2-r^2)+2R\varepsilon_{aij}x_j+2x_ax_i)}
        {g[(t-iR)^2-r^2][(t+iR)^2-r^2]}f(\xi)   \; ,
  \label{solution A}
  \end{equation}
where $r=|\bm{x}|$.
This field configuration clearly has the turning point at $t=0$. The structure
of singularities of this solution is similar to the $\phi^4$ case discussed in
sect.2. In particular, the position of the singularity with smallest
$\mbox{Im}t$ is given by \eq{poles} with $n=0$.

At first sight, the field $A^a_{\mu}$ seems to
violate the asymptotic condition: it does not decay as $|t|\to
\infty$. This is, however, merely a gauge artifact
(one can check that $F_{\mu\nu}$ decays like $1/t$).
Indeed, the part of the solution (\ref{solution A}) that does not go to zero
at $|t|\to\infty$ is concentrated along the light cone and has the asymptotic
form
  \[
  A_0^a = {x^a\over x}  \frac{2R}{\delta^2+R^2}f\left(\frac{1}{2}
   \ln\frac{\delta-iR}{\delta+iR} \right)\; ,
  \]
  \[
  A^a_i = {x^ax^i\over x^2}  \frac{2R}{\delta^2+R^2}f\left(\frac{1}{2}
   \ln\frac{\delta-iR}{\delta+iR} \right)  \; ,
  \]
where $\delta = r\pm t$ depending on whether the initial or final state is
concerned. This non-decaying asymptotics can be killed by the gauge
transformation with the gauge function
\[
  \omega=\cos F(\delta) + i\sigma^an_a\sin F(\delta) \; ,
\]
where $\sigma^a$ are Pauli matrices, $n_a=x_a/r$ and $F(x)$ satisfies the
equation
\[
   {dF\over d\delta}=
   \frac{2R}{\delta^2+R^2}f\left(\frac{1}{2}\ln\frac{\delta-iR}{\delta+iR}
	    \right) \; .
\]
In the new gauge the asymptotics of the transverse spatial components of the
field (\ref{solution A}), which we concentrate on in what follows, reads
\[
  A^a_{\bot i}(x)={2\over gr}\Bigl\{ \sin^2F+(\cos 2F+{\delta\over R} \sin 2F)
           \frac{R^2f(\xi_0)}{\delta^2+R^2}\Bigr\}\varepsilon_{aij}n_j
\]
\[
  +{2\over gr}\Bigl\{ -\frac{1}{2}\sin 2F+(\sin 2F-{\delta\over R}\cos 2F)
  \frac{R^2f(\xi_0)}{\delta^2+R^2}\Bigr\}(\delta_{ai}-n_an_i) \; .
\]
Here $\xi_0=\frac{1}{2}\ln{\delta-iR\over \delta+iR}$.

In the final asymptotic region we find, expanding in $\epsilon$,
\[
  A^a_{\bot i}(x)=\frac{R\epsilon^2}{4gx}\left(
           \frac{R^2(\delta^2-R^2)}{(\delta^2+R^2)^2}\varepsilon_{aij}n_j+
        \frac{2R^3\delta}{(\delta^2+R^2)^2}(\delta_{ai}-n_an_i)\right)\; ,
\]
where $\delta=r-t$.
Performing the spatial Fourier transform and comparing with the plane
wave decomposition
  \[
  A_{\bot i}^a(\bm{k},t) = {(2\pi)^{3/2}\over\sqrt{2\omega_{\bf k}}}
  \Bigl\{
  \epsilon^m_i(\bm{k}) b^{am}_{\bf k} \e^{-i\omega_kt}
  + \epsilon^m_i(-\bm{k}) b^{*am}_{-\bf k} \e^{i\omega_kt}
  \Bigr\} \; ,
  \]
where $\epsilon^m_i(\bm{k})$ are the transverse polarization vectors, we
obtain
  \begin{equation}
  \epsilon^m_{i}(\bm{k}) b^{am}_{\bf k} =
  \epsilon^m_{i}(-\bm{k}) b^{*am}_{-\bf k} =
  {\epsilon^2\sqrt{k\pi}\over 4g}R^2\e^{-kR}\Bigl\{
  -i\varepsilon_{aij}{k_j\over k} + \delta_{ai} - {k_ak_i\over k^2} \Bigr\}\;.
  \label{a[k]:SU[2]}
  \end{equation}
These expressions can be used to calculate the number
particles and the energy of the final state. We find
  \[
  n_f = {3\epsilon^4\pi^2\over 8g^2} \; ,
  \]
  \begin{equation}
  E =  {3\epsilon^4\pi^2\over 4g^2R} \; .
  \label{energy SU2}
  \end{equation}

Consider now the initial state. As in the case of the $\phi^4$ model, we
need to extract only that part of the solution which has singularities
above the contour $\Im\,\delta=T$. The lowest in $\epsilon$
transverse contribution of this type is
  \[
  A^a_{\bot i}(x)=\frac{1}{gx} \frac{3\epsilon^6}{1024}
  \frac{3(\delta/R-i)-i\epsilon^2/8}{[\delta/R-i+i\epsilon^2/8)]^3}
  \ln (\delta/R -i) [ \varepsilon_{aij}n_j - i (\delta_{ai}-n_an_i)]
  \]
  \[
  +\;\mbox{terms regular at Im$\delta>T$}  \; .
  \]
Performing the Fourier transform we obtain
  \[
  \epsilon^m_i f_{\bf k}^{am} =
  -i {3\sqrt{\pi}\epsilon^4\over 64 g \sqrt{k}} \e^{-k(R-T)}
  [\varepsilon_{aij}{k_j\over k} - i(\delta_{ai} - {k_a k_i\over k^2})]
  \int\!du \e^{-k\epsilon^2u/8} {3u-1\over(u+1)^3}
  \]
  \begin{equation}
  \approx -i {3\sqrt{\pi}\epsilon^4\over 64 g \sqrt{k}}
  [\varepsilon_{aij}{k_j\over k} - i(\delta_{ai} - {k_a k_i\over k^2})]
  \e^{-k(R-T)} \; ,
  \label{eps f initial}
  \end{equation}
where the last estimate is obtained at $k\ll 1/\epsilon^2$.
{}From eqs.(\ref{ f_k = a_k }), (\ref{E=aa*})  and (\ref{energy SU2})
we find
  \[
  T-\Delta T = R\Bigl\{ 1- \Bigl({3\epsilon^4\over 256} \Bigr)^{1/3}\Bigr\}
  \;  .
  \]
Making use of this expression we calculate the total number of initial
particles,
  \[
  n_i = {3^{4/3}\pi^2\over 16\cdot 2^{2/3} g^2} \epsilon^{16/3} \; .
  \]
As in the case of the $\phi^4$ model, at small $\epsilon$ the number of
particles in the initial state is much smaller than the number of
particles in the final state. Note also that the characteristic momentum
of the initial particles is $k_{in} \sim \epsilon^{-4/3} \ll
\epsilon^{-2}$, so the assumption of \eq{eps f initial} is justified.

Our next step is to calculate the imaginary part of the action. As in
the case of the $\phi^4$ model, we first evaluate the integral over
$t$ along the contour of fig.2. The only contribution to the
imaginary part comes from the pole which lies between the contour AB
and the real time axis. The remaining integral over $d^3x$ can be done
exactly and gives
  \[
  \Im S = {8\pi^2\over g^2} \; ,
  \]
i.e. the usual instanton action. Estimation of other contributions
into \eq{probability} finally gives
  \begin{equation}
  \sigma = \exp\Bigl\{ -{16\pi^2\over g^2} + O(\epsilon^2) \Bigr\}
  \label{SU2 probability}
  \end{equation}
Again, at small $\epsilon$ the probability of the transition is
exponentially suppressed by the same factor as at zero energy.

Unlike the case of $\phi^4$ model, we can explicitly check that
our solution interpolates between two different topological sectors.
This can be done by calculating the topological charge
along the contour ABCD of fig.1. Since the solution is symmetric
with respect to $t\to -t$, the topological charge evaluated along the real
time axis is zero. Therefore, the only contribution to the topological charge
comes from the pole lying between the contour AB and the real time axis.
However, as one can explicitly check, for our solution the residues of
$(F_{\mu\nu}F^{\mu\nu})$ and $(-F_{\mu\nu}\tilde{F}^{\mu\nu})$ at this pole
coincide, i.e. the solution is anti-self-dual at the singularity. Thus, the
topological charge is

  \[
  Q = - {g^2\over 8\pi^2} \Im S = - 1 \; ,
  \]
and our solution indeed interpolates between different topological
sectors.

\section{Conclusion}

To summarize the results of sects.2 and 3, we have obtained the set of
solutions to the classical field equations that satisfy the requirements
i) and ii) and, therefore, saturate the probability of the instanton-like
transition from some coherent state $\ket{a}$ in the saddle point
approximation. In both $\phi^4$ model and the SU(2) gauge theory the
set of solutions depends on two parameters, the scale parameter $R$ and
the dimensionless parameter $\epsilon$. In the limit of small $\epsilon$
the initial coherent state and the most probable final state for the
corresponding process were found. In both models the number of initial
particles is much less than the number of final particles, while the
latter is much smaller than the characteristic number of particles in the
sphaleron configuration ($1/\lambda$ and $1/g^2$). The characteristic
momenta of particles in the final state are determined by the scale
parameter, $k_f\sim 1/R$, while the characteristic momentum in the initial
state is much larger, $k_i\sim\epsilon^{-4/3}/R$, in both models. Thus, in
the limit $\epsilon\to 0$ any of our solutions describes an
instanton-like transition from a particular coherent state containing
small number of particles at the energy given by eqs.(\ref{energy}) and
(\ref{energy SU2}). The probability of this transition goes to
$\exp(-2S_{inst})$ at $\epsilon\to 0$.

One may ask whether the solutions (\ref{solution}) and (\ref{solution A})
saturate also some transition probability maximized over the initial
states, i.e. the quantity $\sum_{a,b}|\bra{b}SP_EP_{\Lambda}\ket{a}|^2$,
where $P_{\Lambda}$ is the projector onto the eigenspace corresponding to
the eigenvalue $\Lambda$ of the operator $\hat{\Lambda}=\int d\bm{k}
\lambda(\bm{k}) \hat{a}^{\dagger}_{\bf k}\hat{a}_{\bf k}$.  The criterion
is \cite{RST2p} that the ratio $f_{\bf k}/\bar{f}_{\bf k}$ of the
positive- and negative-frequency amplitudes in the initial state is
$\const\cdot\exp(-\lambda_k c)$ where $c$ is some real number. Our
solutions (\ref{solution}) and (\ref{solution A}) do not satisfy this
condition at any real $\lambda_{\bf k}$. The reason is that for both
$\phi^4$ model and SU(2) gauge theory the product $f_{\bf k}\bar{f}_{\bf
k}$ changes the sign at some $k$, in exact analogy to the case of O(3)
sigma-model \cite{RST1}. Therefore, our solutions cannot be interpreted as
saturating some maximum probability and does not automatically set an
upper bound to the two-particle cross section.

To draw some physical conclusions from the existence and properties of the
solutions (\ref{solution}) and (\ref{solution A}), we have to assume that
i) the probability of the transition does not depend, with exponential
accuracy, on the details of the initial state in the limit of small number
of initial particles and ii) there are no other solutions describing the
same process with greater action. If these assumptions
were justified, the two-particle cross section at high energies would
coincide (with exponential accuracy) with the probability of the
transitions described by the solutions (\ref{solution}) and (\ref{solution
A}) in the $\epsilon\to 0$ limit and thus would be suppressed at high
energies by the same suppression factor $\exp(-2S_{inst})$ as at zero
energies.

As discussed in the Introduction, the first assumption is supported by
the perturbative calculations at low energies. Whether the second
assumption is justified or not is presently unclear. This question is
closely related to the (equally unclear) following one: at what energies
the massless theories discussed in sects.2 and 3 can be considered as an
approximations to the massive ones?  Naively, one would expect that the
massless approximation is valid at least at $mR \ll 1$ since $R$ is the
largest scale. In view of eqs.(\ref{energy}) and (\ref{energy SU2}) this
leads to the conditions
  \begin{equation}
  \begin{array}{rl}  \ghost{15pt}
  E\gg \epsilon^2E_{sph} & ~~~ \mbox{in the $\phi^4$ model}\; , \\ \ghost{15pt}
  E\gg \epsilon^4E_{sph} & ~~~ \mbox{in the SU(2) theory}\; ,
  \end{array}
  \label{energy conditions}
  \end{equation}
instead of the expected $E\gg E_{sph}$.

If the conditions (\ref{energy conditions}) are correct, they cause
serious problems with the interpretation of our solutions. Since these
conditions overlap with the region of validity of the perturbation theory
around the zero energy instanton, it is a crucial point to see whether our
results, eqs.(\ref{phi4 probability}) and (\ref{SU2 probability}), can be
reproduced by perturbative methods when we take as an input the initial
states of sects.2 and 3. The perturbative saddle point at small number of
initial particles was studied in ref.\cite{T}. It was found that, in the
leading order approximation, the probability of the transition in that
case grows exponentially with energy and thus differs from the behaviour
we have found in sects.2 and 3. Simple estimates show that for our initial
states there exists another saddle point (in addition to that described in
ref.\cite{T}) with the action and number of particles which parametrically
coincide with those found in sects.2 and 3. If this estimate was reliable,
we would conclude that our saddle point is irrelevant at least at low
energies, since it has smaller action than the saddle point of
ref.\cite{T}.  However, the perturbative expansion about the new saddle
point is ill defined, so that the saddle point itself may be an artifact
of the leading order approximation. Thus, we do not arrive at the
self-consistent picture and cannot make definite conclusions about the
relevance of our solutions to the instanton-like transitions at the
moment.

The authors are indebted to V.A.Rubakov for numerous and helpful discussions.
The work of D.T.S. is supported in part by the Weingart Foundation through a
cooperative agreement with the Department of Physics at UCLA.

%\bibliography{sphal}
%\bibliographystyle{fornp}

\newpage
\begin{picture}(300,200)
\put (50,100){\vector(1,0){200}}
\put (150,40){\vector(0,1){130}}
\thicklines
\put (60,140){\line(1,0){90}}
\put (150,140){\line(0,-1){40}}
\put (150,100){\line(1,0){80}}
\put (60,145){${\bf A}$}
\put (135,145){${\bf B}$}
\put (155,105){${\bf C}$}
\put (230,105){${\bf D}$}
\put (155,135){$iT$}
\put (155,147){$iR$}
\put (148,150){\line(1,0){4}}
\put (155,165){Im~$t$}
\put (240,87){Re~$t$}
\put (140,0){Fig.1}
\end{picture}
\vspace{1cm}

\begin{picture}(300,200)
\put (50,100){\vector(1,0){200}}
\put (150,40){\vector(0,1){130}}
\thicklines
\put (60,140){\line(1,0){90}}
\put (150,140){\line(0,-1){40}}
\put (150,100){\line(1,0){80}}
\put (155,135){$iT$}
\put (155,147){$iR$}
\put (148,150){\line(1,0){4}}
\bezier {120}(145,137)(130,133)(70,133)
\bezier {120}(145,142)(130,145)(70,145)
\bezier {120}(145,145)(130,147)(70,147)
\put (155,165){Im~$t$}
\put (240,87){Re~$t$}
\put (140,0){Fig.2}
\end{picture}

\newpage
\begin{center}
FIGURE CAPTIONS
\end{center}
\begin{enumerate}
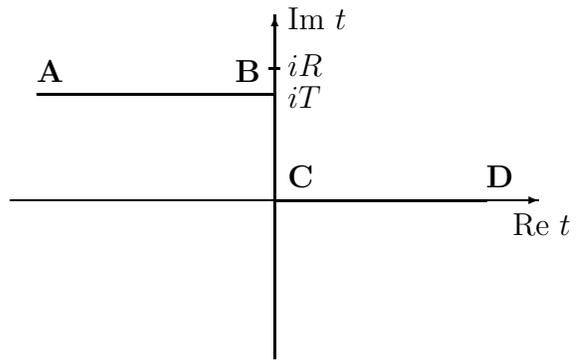
\item[Fig.1]
The contour in the complex time plane where the solution
corresponding to the instanton-like transition is defined. A and D mark
initial and final asymptotic regions, respectively. The line BC represents
the Euclidean part of the evolution.
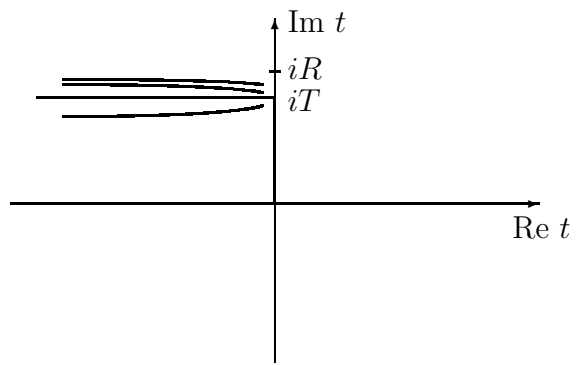
\item[Fig.2] The structure of singularities of the solution in the complex
time plane. Solid curved lines represent the positions of the poles as
$|\bm{x}|$ chagnes from $\infty$ to 0.
\end{enumerate}

\end{document}